\documentclass[conference]{IEEEtran}
\IEEEoverridecommandlockouts

\usepackage{amsmath,amssymb,amsfonts}
\usepackage{algorithmic}
\usepackage{graphicx}
\usepackage{textcomp}
\usepackage{xcolor}

\begin{document}

\title{Optimization on Large Interconnected Graphs and Networks Using Adiabatic Quantum Computation}

\author{\IEEEauthorblockN{Venkat Padmasola}
\IEEEauthorblockA{\textit{Department of Physics} \\
\textit{Center of Quantum Science and Engineering} \\
\textit{Stevens Institute of Technology}\\
Hoboken, N.J., U.S.A. \\
vpadmaso@stevens.edu}
\and
\IEEEauthorblockN{Rupak Chatterjee}
\IEEEauthorblockA{\textit{Department of Physics} \\
\textit{Center of Quantum Science and Engineering} \\
\textit{Stevens Institute of Technology}\\
Hoboken, N.J., U.S.A. \\
rupak.chatterjee@stevens.edu}
}

\maketitle

\begin{abstract}
In this paper, we demonstrate that it is possible to create an adiabatic quantum computing algorithm that solves the shortest path between any two vertices on an undirected graph with at most $3V$ qubits, where $V$ is the number of vertices of the graph. We do so without relying on any classical algorithms, aside from creating an ($V\times V$) adjacency matrix. The objective of this paper is to demonstrate the fact that it is possible to model large graphs on an adiabatic quantum computer using the maximum number of qubits available and random graph generators such as the Barabasi-Albert and the Erdos-Renyi methods which can scale based on a power law.  
\end{abstract}

\section{Introduction}

Adiabatic Quantum Computation (AQC) is a rich field of study in all aspects: fundamental theory, experimental realization, and algorithm development \cite{1}. At its simplest, AQC is a paradigm of computation based on the adiabatic theorem of quantum mechanics which states that a quantum system remains in its instantaneous lowest eigenstate if a given time dependent perturbation is adjusted slowly enough and if there is a gap between the lowest energy eigenvalue and the first excited state. That is, turning on a slowly moving time dependent perturbation to a simple quantum system may change it into a complex system where the ground state energy corresponds to the desired answer of some complex problem. In this definition, a system is composed of many quantum subsystems designed to interact in predetermined configurations. Since the final ground state corresponds to the answer of the problem under investigation, it is the states of these quantum subsystems that contain the needed information, and the slow evolution to the complex quantum system can be thought of as the computation.

Based on this definition for computation, a critical question for the field of AQC is finding quantum system configurations that, when evolved into more complex systems, correspond to difficult classical problems. This is the underlying impetus of developing AQC algorithms. The problems of creating and testing these algorithms has led to the development of the D-Wave Quantum Computer. This quantum hardware relies on the quantum version of the Ising spin model. In this model, quantum spin states are coupled to each other such that the Hamiltonian for $N$-spins reads as follows, 

\begin{equation}
    H = -\sum_{i<j}^{N} J_{ij}\sigma_{i} \sigma_{j}-\sum_{i=1}^{N} h_{i}\sigma_{i}.
\end{equation}

It can be immediately seen that $J_{ij}$ are the spin-spin couplings, $h_{i}$ are the values of the external magnetic fields interacting with site $i$, and $\sigma_{i}$ are the spin states with values at $\pm 1$. In the AQC framework, this will be the final Hamiltonian, where its ground state corresponds to the classical answer. Therefore, the problem of creating an AQC algorithm is to choose appropriate $J_{ij}$ and $h_{i}$ coefficients such that the ground energy state configuration of $\sigma_{i}$ can be translated into the solution to the classical problem under question. It is important to note that this formulation has an equivalent expression known as the Quadratic Unconstrained Binary Optimization (QUBO) problem. In this problem statement the $\sigma_{i}$ become vector components of $\vec{s}$ that take values $0$ or $1$ and the $J_{ij}$ and $h_{i}$ become components of a matrix, $\hat{Q}$, that is upper triangular. Our algorithm will be based on the following QUBO Hamiltonian

\begin{equation}
    H = \vec{s}^{T}\hat{Q}\vec{s}.
\end{equation}

As the name implies, the QUBO method, and by extension the Ising spin model, is very well suited for problems related to optimization. In this case, the lowest energy state would correspond to the optimal solution. A well known optimization problem is finding the shortest path between two vertices in a graph such that the sum of the weights of its constituent edges is minimized. That is, we are interested in the shortest path between any two vertices of a graph, given that their edges correspond to a relevant distance between the points as shown in figure 1.

The problem of solving for the shortest path on a graph has many applications such as delivery routes, social media connections, and online search, \cite{2} In each case, the relevant "vertices" such as stops, people, and websites, respectively, are related by "edges" such as lengths and number of mutual connections. From these examples, it is clear that the problem of finding the shortest path is of great interest.

As such, it follows that this important problem has been studied at both the classical and quantum computational levels. Classically, there are several algorithms that can solve this question and the most popular is the Dijkstra algorithm \cite{3}. This algorithm begins at the starting vertex and locally eliminates adjacent vertices based on their edge values. In this sense, it is known as a greedy algorithm since it creates local minima until it reaches the global minimum. Thus a natural improvement of this algorithm is to find global minimum outright, which is exactly the concept behind AQC. This leads naturally to quantum mechanical formulations to find these shortest paths. A similar problem that may be solved using AQC, being somewhat similar to the shortest path problem, is the traveling salesman problem. This problem is related to our topic as it also is a minimization problem on a graph. The difference lies in the fact that here, we are not concerned with going to each destination, as it might not be relevant for all tasks. 

When these problems are ultimately translated to the quantum computer, the qubit requirement is in some cases proportional to $V^2$, where $V$ is the number of vertices \cite{4}. Thus there is an apparent need to create an algorithm with less hardware strain as scaling the current quantum technologies available is an open problem. In the case where the shortest path was the crux of the algorithm, the constraints depend on the use of a Dijkstra algorithm variant \cite{5}. 

In this paper, we show that it is possible to create an AQC algorithm that solves the shortest path between any two vertices on a graph with at most $3V$ qubits, where $V$ is the number of vertices. We also do so without relying on any classical algorithms, aside from the creating an ($V\times V$) adjacency matrix. In section II, we define the problem statement and several sub-problems related to creating the QUBO equation. Afterwards, we solve these problems and suggest how matrices related to a given graph can be used to formulate the QUBO equation. In section III, we present findings for a randomly generated graph with eight vertices and varied edges. Since this graph is small and serves to test our algorithm, we first analyze classical solutions to a problem with $2^8$ qubit combinations. We thereafter present our results found by directly evaluating the QUBO equation on the D-Wave 2000Q processor. In section IV, we extend our methodology to larger graphs created via the Barabasi-Albert and Erdos- Renyi models. These models address current issues with embedding large graphs on the D-Wave architecture. In section V, our quantum algorithm is compared to several classical shortest-path algorithms. The results are given for varied edge probabilities and total vertices. This results in a comparison of time and space complexity for our algorithm versus the chosen classical algorithms. In section VI, we previously address issues with the time complexity analysis that arise from the annealing schedule of AQC. Lastly, in section VII, we summarize our key findings and present potential future improvements.

\section{Methodology}

When formulating an AQC algorithm in the form of a QUBO problem, as in equation (2), the matrix $\hat{Q}$ that defines the problem is upper triangular and therefore, the only effects on the final QUBO problem are due to diagonal elements and upper off-diagonal elements. Upon carrying out the matrix-vector multiplication, it is apparent that diagonal elements correspond to qubit biases and off-diagonal elements act as qubit-qubit coupling terms. Fortunately, graphs have been studied at great length and the study of graphs on computers has led to the creation of many matrices that describe their features. For our purposes, the starting point will be an adjacency matrix. An adjacency matrix, $\hat{A}$, for an $V$-vertex graph, is defined as an $V\times V$ matrix with elements $A_{ij}$ equal to one if vertices $i$ and $j$ share an edge and zero otherwise. The definition ensures that the matrix is symmetric therefore no information is lost when we perform an upper triangularization. Naively using the adjacency matrix as the QUBO problem matrix will result in the wrong solution. We need to explicitly define edge-weights to rectify this and limit the edgeweights to 3 at most for accuracy in the solutions. We do this by ensuring that our QUBO formulation yields a symmetric matrix which can be diagonalised. We can achieve this by adding the inverse-adjacency matrix where the 1's are converted to 0's and vice-versa into our QUBO equation along with the identity matrix and the adjacency matrix as show below in equation (3). The edgeweights are tuned by introducing scalar coefficients to the three matrices described above, which completes our QUBO equation. Graphs with a more diverse set of edgeweights prove to be more problematic to run on the QPU as a QUBO model. In the model, the vertices directly correspond to the qubits after which we translate a physical graph into the QUBO format.

 A Graph is initialised G = (V, E) such that there are V vertices, and an initial connectivity defined by E which has all the values for the edges present in the graph via a V by V square matrix. The entries of this graph matrix will all be 0 or 1 based on whether there is a connection between 2 vertices or not respectively. Once we create the QUBO model from this graph, we observe that the accuracy in estimating the most optimal solution is obtained if the edge values are constrained to 0, 1, 2 and 3. Increasing the range of values for edge weights can be computationally tedious for estimation of the classical counterparts. To further define our problem statement, the Graph G = (V, E ) can be broken down into subsets of vertices and edges such that 
 \begin{equation}
 V = \begin{pmatrix}
 v_1\\ 
 v_2 \\
 v_3\\ 
 .\\
 .\\
 .\\ 
 v_n
 \end{pmatrix}, \;\;\; E = \begin{pmatrix}
e_{1_1} & e_{1_2} & e_{1_3} & ...... & e_{1_n}\\
e_{2_1} & e_{2_2} & e_{2_3} & ...... & e_{2_n}\\
.\\
.\\
.\\
.\\
e_{n_1} & e_{n_2} & e_{n_3} & ...... & e_{n_n}\\ 
\end{pmatrix}
\end{equation}

\subsection{The QUBO Equation}

In the previous section, the three necessary components of the algorithm have been determined. In this section, they are put together to create the QUBO problem. The QUBO problem is then given an exact form related to the randomized graphs that were solved to verify the accuracy of the results. The adjacency matrix will be defined as $\hat{A}$, the identity matrix as $\hat{I}$, and the inverted adjacency matrix as $\hat{N}$.

\begin{equation}
   \hat{Q} = -\alpha \hat{A} + \beta \hat{I} + \gamma \hat{N} 
\end{equation}

In equation (4), $\hat{Q}$ is the QUBO problem matrix before upper triangularization and $\alpha,\beta,\gamma$ are positive scalars that are determined by features of the original graph. The QUBO problem is created by upper triangularization, defined by $UT$, of $\hat{Q}$, multiplication by $\vec{s}$, $\vec{s}^T$ and subtraction of terms related to $s_{start}$ and $s_{end}$,

\begin{equation}
     H(\vec{s}) = \vec{s}^{T}UT[\hat{Q}]\vec{s} -\delta(s_{start}+s_{end}+s_{start}s_{end})
\end{equation}
From equation (5), $H(\vec{s})$ is the function corresponding to the QUBO problem. The question of the shortest path between two vertices on a graph becomes finding sufficient values for $\alpha,\beta,\gamma$, and $\delta$. 

\section{Results for "Small" Graphs}

In this section, we detail the graphs we randomly chose for verifying the accuracy of the results. The graphs are all 8 vertices in size to ensure ease of classical verification. The edge number is varied to be 10 and 12, for simplicity we have left all edges to have an equal "length" of 1. The routine is simplified on the D-Wave computer by making the start vertex 1 and the end vertex 8 in each case. Since we have picked an 8 vertex graph, the adjacency matrix is an 8$\times$8 matrix and the total number of distinct path combinations with be $2^8$ which is 256 path combinations in total. The energy per path combination is computed and the energies are presented as a discrete plot. These plots highlight the complexity of finding a global minimum, which the QPU is well equipped to handle. Furthermore, Graph \#2 has the interesting feature of having 2 distinct minimal paths. The previous methodology made no distinction between the shortest paths, so we expect the D-Wave to return each path with equal likelihood. Upon preliminary calculations we have found that choosing $\alpha = 1$, $\beta = 1$, $\gamma = 2$, and $\delta = 3$ sufficient to ensure the shortest path corresponds to the lowest energy. Section III.1 will be dedicated to the classical computation done to verify the QUBO problem before evaluation in the D-Wave hardware. The D-Wave results, which exactly match the results obtained via classical computation are presented below.

\subsection{Verification: Classical Computation}

As detailed in Section II, the verification of the classical computation is easily accomplished by determining the graph’s adjacency matrix and then calculating the QUBO functions value for each qubit permutation. Even for the test graphs, this means $2^8$ or $256$ unique computations. The results needed to showcase that the shortest path would then be quite difficult to display properly. To circumvent this issue, the qubit outputs have been converted into binary. This conversion allows the QUBO values to plotted instead of listed. For each graph, we highlight the binary digit corresponding to the shortest path with a gray line.

The graph below corresponds to Graph \#1 in figure 2. Clearly the shortest path on this graph is $1 \leftrightarrow 7 \leftrightarrow 8$, or qubits 1, 7, and 8 valued at 1 while the rest are 0. This combination is shown on the graph and occurs at a value of $-6$. All other qubit permutations clearly result in larger QUBO function values. 

\begin{figure}[h!]
\includegraphics[width=8cm]{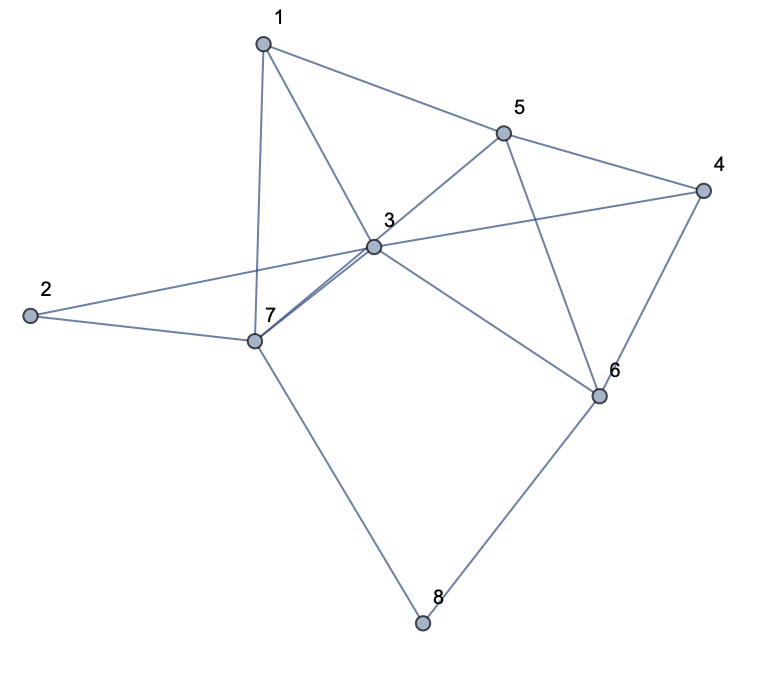}
\centering
\caption{Graph \#1 consists of 8 vertices and 14 edges. The vertex numbering directly corresponds to the qubit numbering on the D-Wave computer. Each edge has "length" of 1 despite the varied image length needed to display the graph.}
\end{figure}

\begin{figure}[h!]
\includegraphics[width=9cm]{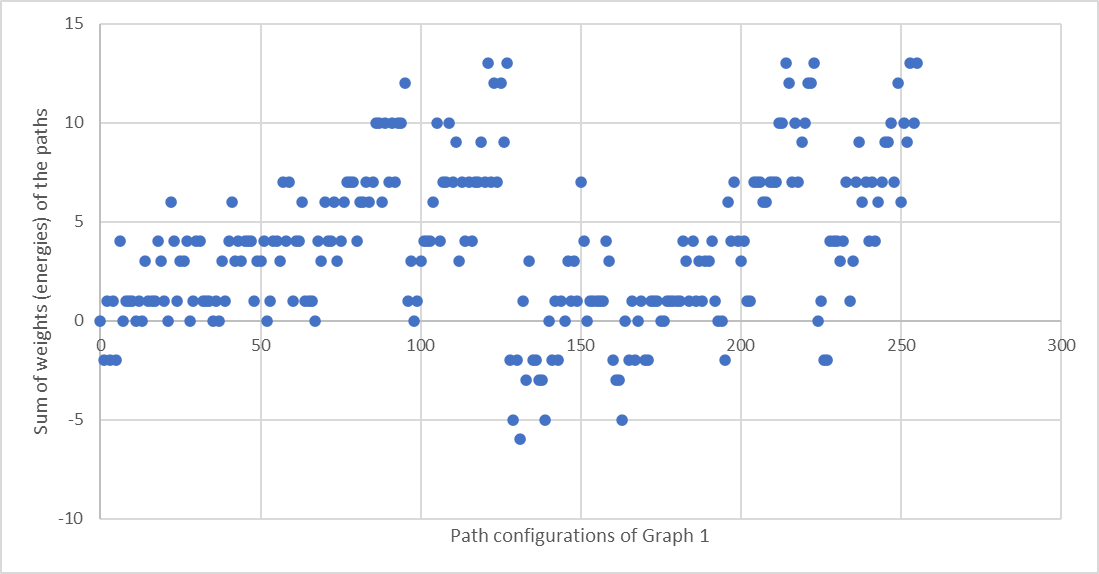}
\centering
\caption{Results of evaluating Equation 4 with $\alpha = 1$, $\beta = 1$, $\gamma = 2$, and $\delta = 3$ for Graph \#1. The binary value of $1 \leftrightarrow 7 \leftrightarrow 8$ has the global minimum of $-6$, showcasing the algorithm should converge to the correct value.}
\end{figure}

Next, we solve the shortest path for Graph \#2 using the algorithm. Unlike Graph \#1, there are two solutions that are considered the shortest path. The below graph showcases the QUBO function values where the two shortest paths, $1 \leftrightarrow 2 \leftrightarrow 8$ and $1 \leftrightarrow 3 \leftrightarrow 8$, are highlighted by gray lines. It is interesting to note that the algorithm gives both paths the same lowest value of $-6$.

\begin{figure}[h!]
\includegraphics[width=8cm]{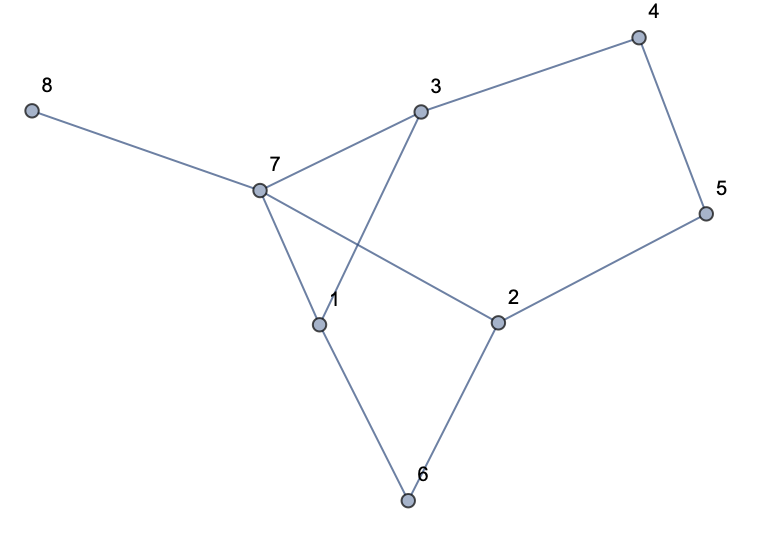}
\centering
\caption{Graph \#2 consists of 8 vertices and 12 edges. The vertex numbering directly corresponds to the qubit numbering on the D-Wave computer. Each edge has "length" of 1 despite the varied image length needed to display the graph.}
\end{figure}

\begin{figure}[h!]
\includegraphics[width=9cm]{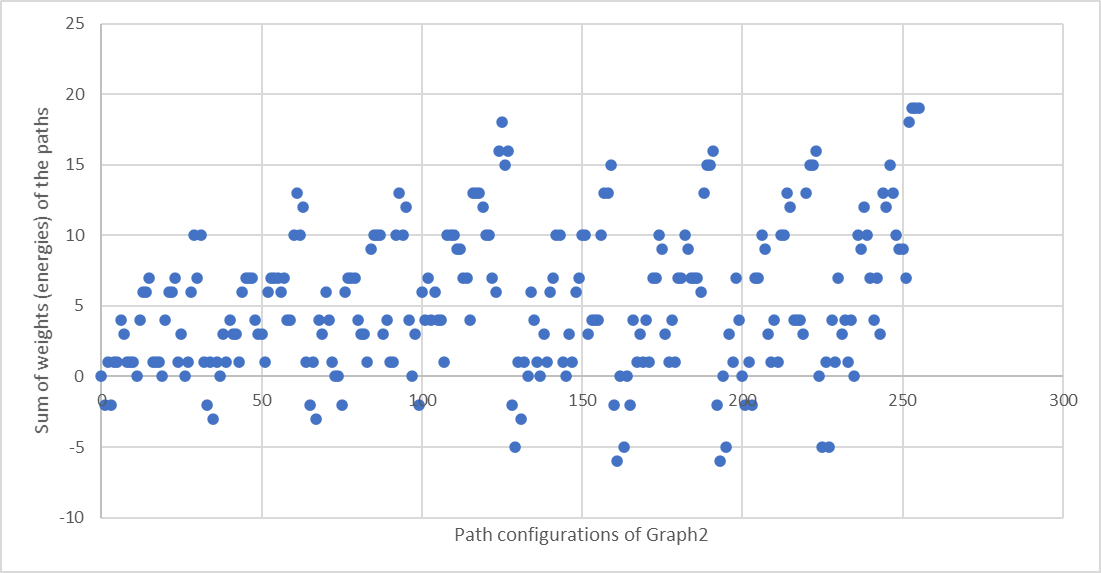}
\centering
\caption{Results of evaluating Equation 4 with $\alpha = 1$, $\beta = 1$, $\gamma = 2$, and $\delta = 3$ for Graph \#2. The binary values of $1 \leftrightarrow 2 \leftrightarrow 8$ and $1 \leftrightarrow 3 \leftrightarrow 8$ have the global minimum of $-6$. This showcases the algorithm can also find multiple shortest paths with equal likelihood.}
\end{figure}

In all the test cases, it is apparent that the only global minimum of these QUBO functions occur when the qubits reflect the shortest path. Also, there are a few paths with the next lowest value of $-5$. In all cases, the solutions either reflect only vertices 1 and 8 or a path with an additional step. The first case is due to the potentially heavy-handed $\delta$, ensuring the start and end vertex are included in the solutions. The second case confirms that the objective function corresponds to minimizing physical length. Furthermore, the maximum value of the QUBO values occurs when all the vertices are on. This ensures the D-Wave will never return something that is clearly wrong.

\subsection{The graph optimisation: D-Wave Results for small test graphs}

The sampling has been set to read 1000 outputs generated by running the program on the D-Wave system. The results for three test graphs (\#1, \#2) converted into QUBO problems are displayed below.

\begin{table}[h!]
\centering
\begin{tabular}{ |c|c|c| } 
\hline
Energy Value & Solution Path & \# of Occurrences \\
\hline
-6 &  $1 \leftrightarrow 7 \leftrightarrow 8$ & 991\\
-5 &  $1 \leftrightarrow 8$ & 4\\
-5 &  $1 \leftrightarrow 3 \leftrightarrow 7 \leftrightarrow 8$ & 3\\
-5 & $1 \leftrightarrow 5 \leftrightarrow 7 \leftrightarrow 8$ & 1\\
\hline
\end{tabular}
\caption{Experimental results of the shortest path for Graph \#1 on the D-Wave computer. The table shows the paths returned with the most frequency after 1000 runs. The correct solution is returned with overwhelmingly frequency.}

\end{table}

\begin{figure}[h!]
\includegraphics[width=9.0cm]{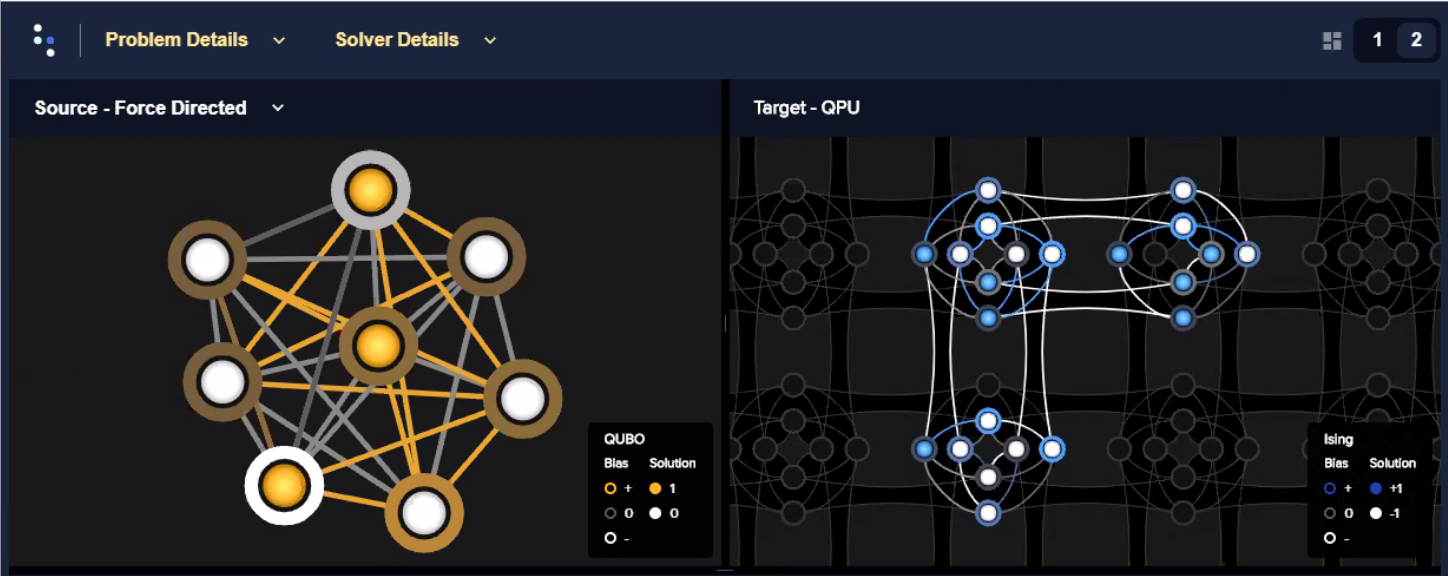}
\centering
\caption{Graphic representation of the QUBO problem and target QPU on the D-Wave computer for Graph \#1. The biases, and QUBO and QPU parameters are outlined in the figure legend.}
\end{figure}

\begin{table}[h!]
\centering
\begin{tabular}{ |c|c|c| } 
\hline
Energy Value & Solution Path & \# of Occurrences \\
\hline
-6 &  $1 \leftrightarrow 2 \leftrightarrow 8$ & 562\\
-5 &  $1 \leftrightarrow 3 \leftrightarrow 8$ & 418\\
-5 &  $1 \leftrightarrow 2 \leftrightarrow 3 \leftrightarrow 8$ & 7\\
-5 &  $1 \leftrightarrow 2 \leftrightarrow 3 \leftrightarrow 7 \leftrightarrow 8$ & 7\\
-5 &  $1 \leftrightarrow 2 \leftrightarrow 7 \leftrightarrow 8$ & 2\\
\hline
\end{tabular}
\caption{Experimental results of the shortest path for Graph \#2 on the D-Wave computer. The table shows the paths returned with the most frequency after 1000 runs. Despite having multiple shortest paths, each are returned with high frequency. This showcases our algorithm has no bias for a particular shortest path.}
\end{table}

\begin{figure}
\includegraphics[width=9.0cm]{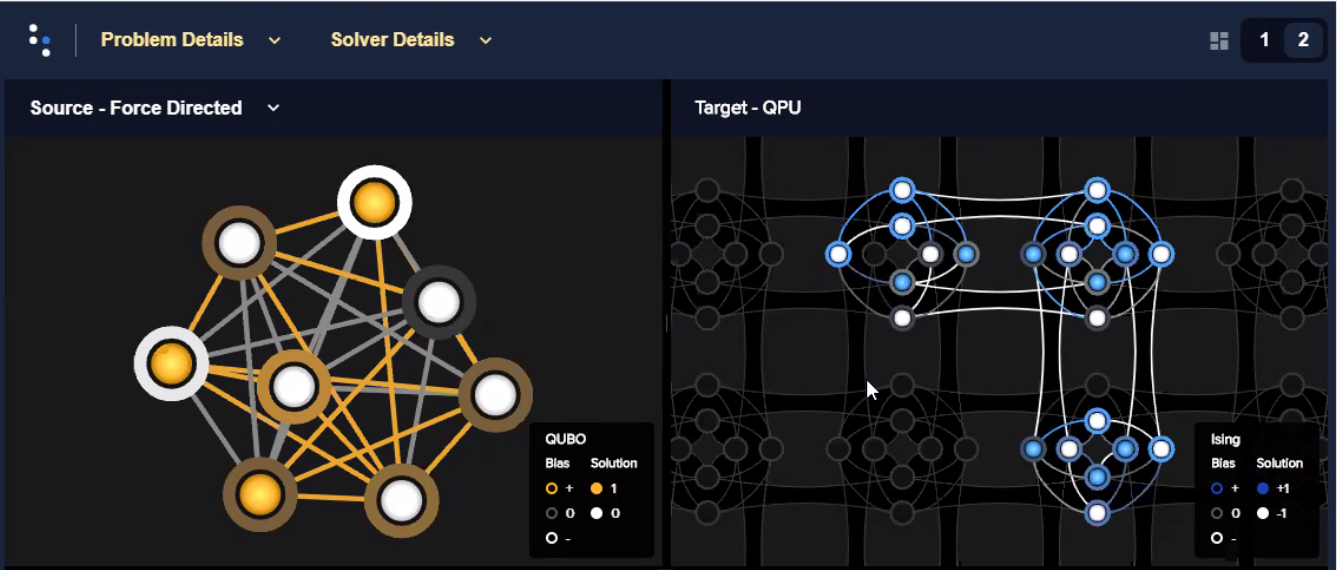}
\centering
\caption{Graphic representation of the QUBO problem and target QPU on the D-Wave computer for Graph \# 2. The biases, and QUBO and QPU parameters are outlined in the figure legend. }
\end{figure}

The vertices on the Source Graphs, the left graph in figures 8, 9, and 10, represent the QUBO parameters, $s_{i}$, and the edges between them are the coefficients in the QUBO matrix, $\hat{Q}$. The target QPU graphs in the mentioned figures represent the embedding of the QUBO graph onto the D-Wave chimera graph architecture. In this case, the vertices represent the physical qubits and the edges between them are defined as the couplings between the qubits. The QPU is submitted to the D-Wave computer and the results are generated in a single step over the energy landscape of the quantum processor. The QPU system is initialized as an equal superposition of each qubit. Afterward, the QPU has been evolved in time adiabatically and the results are recorded. The process is repeated for 1000 runs. As expected the lowest energy configuration is the most read measurement in the series of 1000 measurements on the QPU, which matches the value obtained in classical simulations.

\section{Annealing Time}
We have noticed that there was an initial time delay observed in the plot of the programming time on the Quantum annealer. This is due to the fact that when the initial Hamiltonian is presented, according to the Adiabatic theorem, the system evolves over time until it stabilizes into a final Hamiltonian, from which we can track the ground state energy which is the lowest energy state over the energy landscape.
 
 \begin{equation}
H(s) = (1 - s)H_{S} + sH_{P}
 \end{equation}
From the time dependent Schrodinger we obtain the adiabatic evolution of the initial Hamiltonian to the final Hamiltonian.
 
\begin{equation}
-i\hbar\frac{d\psi}{dt} = H\psi
\end{equation}

The minimum time $\tau$ to anneal is given as
\begin{equation}
\tau = -\int_{0}^{1} \left(\dfrac{dH(s)}{ds}\dfrac{1}{\gamma(s)^2}\right)ds
\end{equation}
 where $\gamma$ is the minimum spectral gap, which is the energy difference between the ground state and the first excited state in the energy spectrum. It is hard to extrapolate exact values and bounds for $\gamma$ and $\tau$ since we are dealing with very large graphs. The annealing time range for the chimera processor is from 1 to 2000 microseconds. The total sampling time in the QPU is usually bounded by and proportional to the annealing time multiplied by the total number of samples. These are default solver parameters that can be set to custom values. The extra time taken to run the problem depends on the graph embedding, the post-processing and readout time.
 
 Figure 7 gives us insight into how a problem is run on a quantum annealer. We collected the runtime data for different anneal times and have restricted the number of samples per job to be 10 in order for the time range to be comparable to the Dijkstra algorithm which is quite efficient by itself. The plot itself has been scaled to a timescale which is of the order of $10^{-7}$ seconds.
 
 \begin{figure}[h!]
\centering
\includegraphics[width=1.1\linewidth]{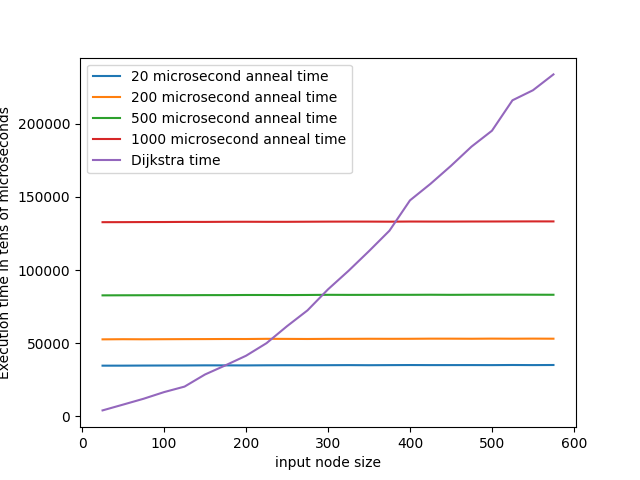}
\caption{Comparison of the different runtimes as a function of total number of samples and anneal times, estimating their runtime Vs. Input size}
\label{"vertices vs time"}
\end{figure}
 
The quantum processing times appear to be linear in these plots but that is not the case, The individual plots of the time vs size for the anneal times of 20, 200, 500 and a 1000 microseconds have been generated and are presented in figures 8, 9, 10 and 11.
 
  \begin{figure}[h!]
\centering
\includegraphics[width=1.1\linewidth]{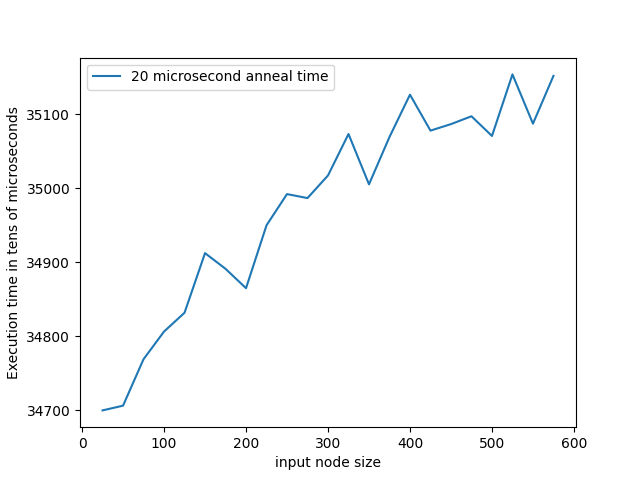}
\caption{Vertices vs time for a 20 microsecond anneal time}
\label{"vertices vs time for a 20 microsecond anneal time"}
\end{figure}

  \begin{figure}[h!]
\centering
\includegraphics[width=1.1\linewidth]{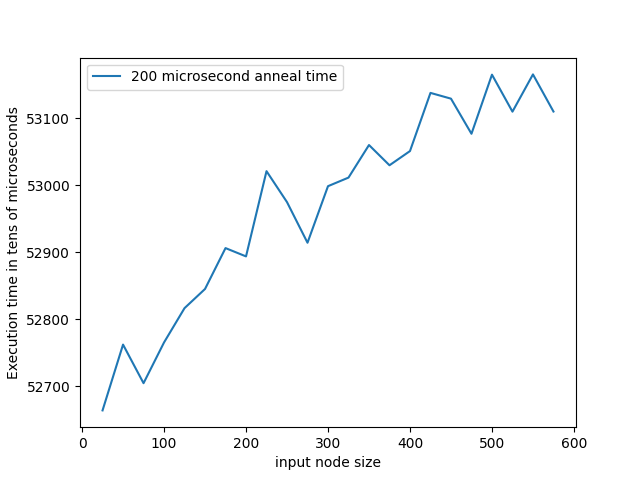}
\caption{Vertices vs time for a 200 microsecond anneal time}
\label{"vertices vs time for a 200 microsecond anneal time"}
\end{figure}

  \begin{figure}[h!]
\centering
\includegraphics[width=1.1\linewidth]{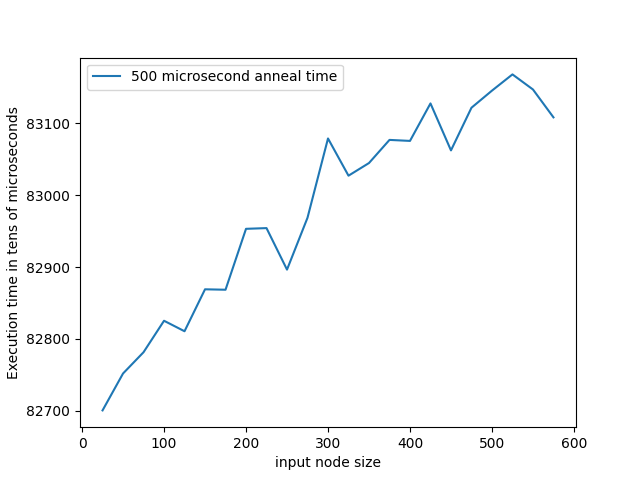}
\caption{Vertices vs time for a 500 microsecond anneal time}
\label{"vertices vs time for a 500 microsecond anneal time"}
\end{figure}

  \begin{figure}[h!]
\centering
\includegraphics[width=1.1\linewidth]{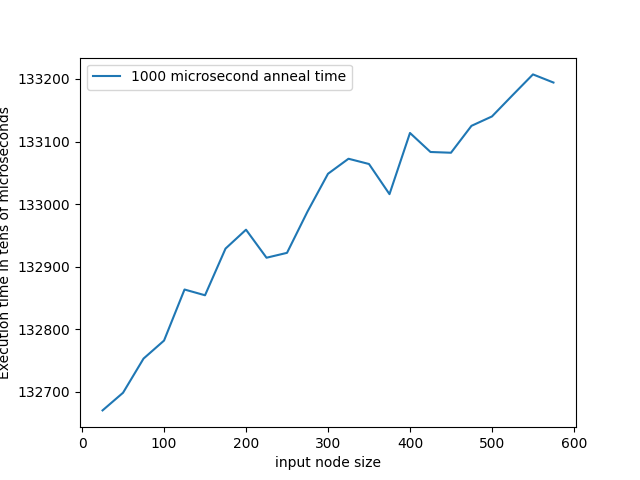}
\caption{Vertices vs time for a 1000 microsecond anneal time}
\label{"vertices vs time for a 1000 microsecond anneal time"}
\end{figure}
 
 \section{Creating Qubo models for large Graphs}
 \subsection{The objective function formulated as an Ising Hamiltonian}
 Although we were able to demonstrate that the QPU can initialise large graphs. One major criticism would be that the degree of connectivity in the input graph has to be limited to at most 5 connections per vertex to ensure the accuracy in the sampling. This limitation can be overcome as the technology matures in the future.
 
 We then applied the Dijkstra, Floyd Warshall and the Bellman ford algorithms to find the shortest paths given the biases on the vertices and couplings between them as the linear and quadratic terms of the Binary quadratic model.
 The Hamiltonian operator given by (1) operates on the qubits to give a scalar observable value according the time-independent Schrodinger equation, $H\psi = E\psi$ where $\psi$ represents the multi-qubit wavefunction. To obtain the physical observable value of E, from statistical analysis we determine all the transition energies of this multi-qubit system as a 2-D energy surface.
\begin{equation}
\langle H \rangle =\sum_{j=1}^N\sum_{i=1}^{N}\int\psi_i^*H\psi_j d\tau = E
\end{equation}

This equation is then mapped to the QUBO equation shown below.

\begin{equation*}
    \hat{Q} = -\alpha \hat{A} + \beta \hat{I} + \gamma \hat{N}
 \end{equation*}

We do not need to explicitly define edgeweights since the parameters $\alpha$, $\beta$ and $\gamma$ make sure the adjacency matrix A is converted into the QUBO matrix which is essentially a cost matrix which has the form 

\begin{equation}
C = \begin{pmatrix}
c_{11} & c_{12} & c_{13} & ...... & c_{1n}\\
c_{21} & c_{22} & c_{23} & ...... & c_{2n}\\
.\\
.\\
.\\
.\\
c_{n1} & c_{n2} & c_{n3} & ...... & c_{nn}\\ 
\end{pmatrix}
\end{equation}

 \subsection{Embedding and Hardware Topology}
 The D-wave processor is modelled from the chimera topology, and the embedding from a source graph are mapped on the the chimera using D-wave's embedding composite. For each vertex there is a maximum of three qubits initialised, where 2 qubits are in superposition and one qubit stores the information about the state of the entangled qubits. The actual physical qubits present on the D-wave chip are RF SQUID flux qubits \cite{6}. The embedding is defined as an isomorphism between 2 vector spaces where one space represents the vectors suitable to be solutions for the Ising model and the other represents the physical state of the qubits on the processor.
 
 We define an edge weight function $f$ which is to be minimised such that
 \begin{equation*}
Shortest-path = min[\sum_{i=1}^{n-1} f(c_{i \;i+1})].
 \end{equation*}
When applied to solving the shortest path as applied on a 2-D map using Euclidean coordinates. The shortest path problem can be understood as a distance matrix with edge weights, which can be represented as the product of two square matrices $X$ and $Y$ of size $N$ such that $X*Y = Q$ and $Q_{ij}$ = $\sum_{1}^{j}min(X_{ik} + Y_{ik})$ which is a min-plus algebraic formulation. The product of the matrices should be symmetric, to ensure that the resulting matrix is self adjoint which is essential to reinforce the fact that the QUBO matrix should be Hermitian. Assuming we are operating in euclidean space, the vertices of the graph correspond to coordinates $x_i$ and $x_j$ and the elements of our QUBO matrix should correspond to the distance $d_{ij}$ such that
 
 \begin{equation*}
    d_{ij} = ||x_i - x_j||^2
\end{equation*}
 
Since our formulation requires an identity matrix and a non-physical matrix which is a binary inverse of the adjacency matrix

\begin{equation}
\begin{array}{c}
     H(\vec{s}) = \vec{s}^{T}[-\alpha \hat{A} + \beta \hat{I} + \gamma \hat{N}]\vec{s}  \\ \\
     -\delta(s_{start}+s_{end}+s_{start}s_{end})
\end{array}
\end{equation}
 We need only to change the values of the scalar coefficients $\alpha$, $\beta$, $\gamma$, $\delta$, to these matrices as we increase the graph size. If the start and end vertices are unspecified we can effectively remove the parameter $\delta$. However adding new vertices and defining the connections between vertices becomes more and more tedious as the size of the graph increases if we have to scale the graph without manually adding vertices and defining the connectivity, we need to introduce a graph generator method such that the network grows in a simple manner obeying a certain mathematical relation which is, they should be scale free to simulate the simplest possible case for handling large graphs.  Since we only need the graph and coefficients as an input to generate $Q$, we can fine-tune our abstract problem by extracting the sub-matrices of $Q$, which can in themselves be initialised as a QUBO problem. 
 
 The qubits in the D-Wave processor are made of superconducting niobium rings, fabricated to form Josephson junctions \cite{7}, which are arranged on the QPU in the form of arrays of Chimera unit cells with each unit cell consisting of 8 qubits. As of now, the D-wave 2000Q has a maximum of 2048 qubits on a single QPU, with almost 5 connections per qubit.

\subsection{Scale free Networks}
After confirming that the test graphs produce accurate results, we now simulate large graphs on the D-wave QPU and perform a comparative analysis to all the pre-existing shortest path algorithms. Large graphs are defined by high degree distribution and large vertex count. As expected, the quantum annealing method gives out all possible paths simultaneously in a time scale, which is at least faster by a factor of 10 as compared to the most efficient pre-existing classical algorithms to determine the shortest paths between 2 points in a densely populated and interconnected network. 
 
 Some of the input graphs are initialised using common graph generators for scale-free networks which are simulated from real world graph structures such as fluctuating stock markets, massive social networks. As such scale free networks are characterised by 2 main factors which are growth and preferential attachment. growth refers to the fact that the number of vertices in the network increases in time. Preferential attachment refers to the fact that new vertices tend to connect more to existing vertices with large degree.  This is the major reason why we have used Barabasi-Albert and Erdos-Renyi models to generate graphs. For an approximately scale free network, its degree distribution follows a power law such that 
 \begin{equation}
  p(k) = k^{-\gamma}                
 \end{equation}
 
 To generate the preferential attachment, we introduce a mathematical relation which determines how a new vertex connects to an existing vertex where the probability of a new vertex connecting to an existing vertex (probability of acquiring an edge) is given by
 \begin{equation}
 p_i = \dfrac{k_i}{\sum_{j}k_j}
 \end{equation}
 
 Most scale free networks have a $\gamma$ value of either 2 or 3(Add citation to Network Science-Barabasi). For the Barabasi-Albert model, to a starting set of connected vertices $n_0$, new vertices with n edges are added in a timely manner such that they are connected to n different pre-existing vertices. The rate at which a vertex obtains edges is equal to the number of edges added times the probability of acquiring an edge as shown above
 
 \begin{equation}
 \frac{dk_i}{dt} = np_i = n\left( \dfrac{k_i}{\sum_{j}k_j} \right)
 \end{equation}
 
 This is a first order differential equation which can be solved in a simple manner and has a $\gamma$ value of 3 and a degree distribution of the form,
 
 \begin{equation}
 \frac{d}{dk}P(k_i < k) = \frac{2m^2 t}{k^3} \frac{1}{(m_0 + t)}
 \end{equation}
 
 The degree distribution provides insight into the structure of the networks. The Barabasi-Albert network in particular is an interesting choice when modelling complex networks due to the fact that highly connected vertices are more probable to acquire more connections as new vertices are added to the network. It becomes apparent that with increasing value of n, the degree distribution of a generated network diversifies. With greater complexity in a network, the chances of having a wide range of degrees also increases. Intuitively, the Barabasi-Albert generator comes very close to mimicking these traits, which allows us to build models for large networks with little effort. To demonstrate this, the Degree distributions of Barabasi-Albert graphs with n having values of 1, 5, 20 are plotted in the figure below.

\begin{figure}
\centering
\includegraphics[width=1.0\linewidth]{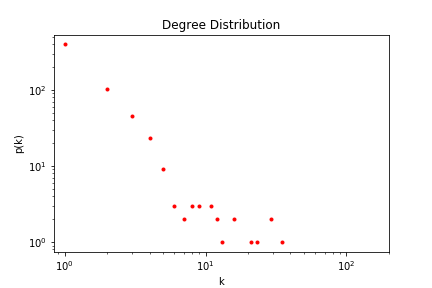}
\caption{Degree distribution of a Barabasi-Albert random graph with n = 1 and having 600 vertices. P(k) is scaled to percentage values}
\label{"a"}
\end{figure}

\begin{figure}
\centering
\includegraphics[width=1.0\linewidth]{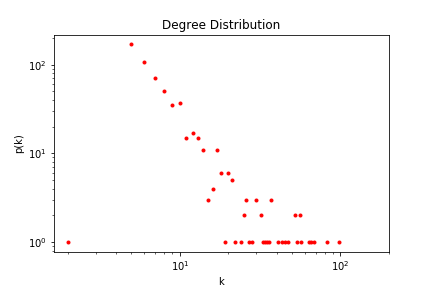}
\caption{Degree distribution of a Barabasi-Albert random graph with n = 5 and having 600 vertices. P(k) is scaled to percentage values}
\label{"b"}
\end{figure}

\begin{figure}
\centering
\includegraphics[width=1.0\linewidth]{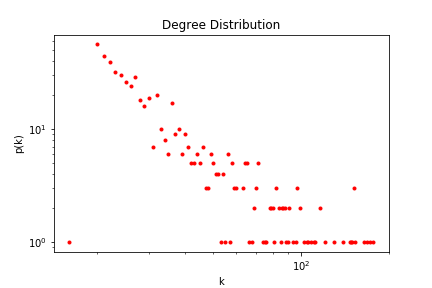}
\caption{Degree distribution of a Barabasi-Albert random graph with n = 20 and having 600 vertices. P(k) is scaled to percentage values}
\label{"c"}
\end{figure}

 The Erdos-Renyi model on the other hand, follows an attachment behavior based on the binomial distribution. We choose v vertices from among $v_0$ in $v_0\choose v$ ways, and $p^v$ is the probability that they will have edges to v vertices, the probability that the rest of the $v_0$ - v vertices do not have an edge is given by (1 - $p^{v_0 - v}$). It is observed for Erdos-Renyi networks that if the number of vertices becomes large the attachment behavior follows the poisson distribution.

The degree distribution for a small Erdos-Renyi graph with 600 vertices follows the equation below.
 \begin{equation}
 Probability(v) = {v_0\choose v} p^v (1 - p^{v_0 - v})  
 \end{equation}
 
\begin{figure}[h!]
\includegraphics[width=9.0cm]{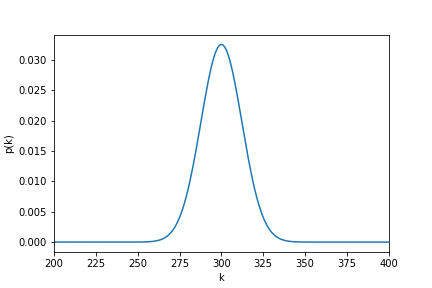}
\centering
\caption{Degree distribution of Erdos-Renyi random graph with 600 vertices.}
\end{figure}

 For demonstration purposes, we have selected a network modelled on the internet, which has both a large input size a good connectivity and relatively scale free following a preferential attachment similar to the Barabasi-Albert graph. Given the input parameters such as the edgeweights and the weights corresponding to each vertex, the program returns all the solutions, sampled over an energy landscape and then we sort this data, to obtain an ordered list from which we can identify the shortest paths. For convenience sake in generating such large graphs we make sure that the graph is undirected and the weights of the edges are usually 0, 1, 2 or 3 at most.
 
 We have verified that the program that was written for the D-Wave QPU, is able to model a wide variety of complex networks, and return their corresponding Binary Quadratic models, which can then be sampled for results. However, we have also observed that for efficient embedding of the problems on the QPU, we have to select input sizes in such a way that the number of available qubits should be at least 3 times the size of the input of the graph.
 
Once the program is executed on the D-wave QPU for a 1000 samples, the problem inspector tool gives all the details of computation time, any chain breaks, details of broken vertices and a fault analysis of broken constraints in the constraint satisfaction step which is optionally inputted if needed.

It can be seen here that most of the qubits which are available on the chip are being utilised for the input graph with 600 vertices, which verifies the statement that we need our initialised qubits to be at least 3 times more in number as compared to the number of input vertices. The D-wave problem inspector tool provides a very convenient way to visualise the graph isomorphism that occurs between 2 Graph structures, which in itself is an NP-hard problem. This inherently occurs when we define the embedding of our Graph on to the QPU, and thereby it becomes important to carefully determine, what constraints need to be kept in mind before initialising any graph structure. One such constraint is that the number of connections per vertex is limited to the connectivity of the Chimera graph. This seems to be a major obstacle in solving more complex optimisation problems.

\begin{figure}
\centering
\includegraphics[width=.7\linewidth]{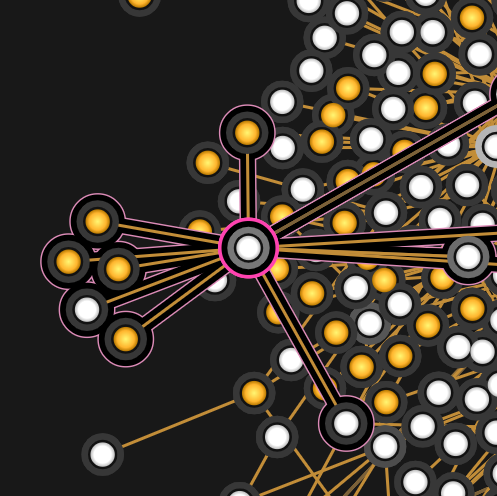}
\caption{Visualising a single vertex in the graph structure}
\label{"a"}
\end{figure}

\begin{figure}
\centering
\includegraphics[width=.7\linewidth]{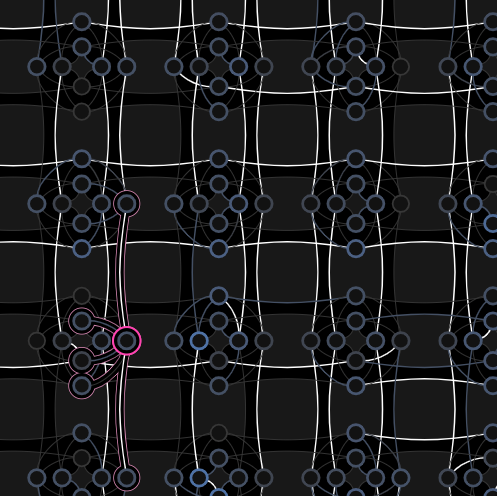}
\caption{Micro-embedding of the vertices in the QPU}
\label{"b"}
\end{figure}

\section{Results}
We tested the shortest path problem both on the D-wave 2000Q machine and a computer with an Intel Processor QuadCore i5-8250U CPU @ 1.60GHz-1.80 GHz clock speed and 8 GB of RAM, which was used to run the Dijkstra, Bellman-Ford and Floyd-Warshall algorithms. The run time was recorded and graphs were plotted with increasing input size. We have noticed that embedding large graphs in a single instance has been problematic as opposed to embedding small graphs and sequentially expanding them by adding more vertices and connections to the greatest extent possible with respect to the size. In order to successfully do this, we have used sparsely populated Barabasi-Albert graphs with atmost 2 connections per vertex to ensure easier embedding on the Dwave processor. We have also manually recorded the time it took for the send the problem to the Dwave machine remotely via the cloud and return the solution to test the utility of quantum annealers in their current state.

\subsection{Classical Algorithms Analysis and performance of the QPU with increasing input sizes}    
A detailed analysis of the different shortest path algorithms have been done, and the computation time has been recorded in the table below. We have made a comparison of the time it takes for the standard classical shortest path algorithms which is demonstrated in figure 15, such as the Dijkstra, Bellman-Ford and the Floyd-Warshall algorithms, to the time it takes for a solution on the D-wave QPU for a graph with uniform edge weights. For this analysis, the parameters for the D-wave machine have been set in a way to give us the best possible run times. The program has been tested for different input sizes and the estimated programming time for a hundred samples has been generated and shown in the table below, figure 14 gives us information about the variation of the total quantum annealing time with respect of the size of the graph to be embedded. For the classical algorithms we have used the built in shortest path algorithms from the networkx graph library.
\begin{table}[ht!]
\centering
\scalebox{0.85}{
\begin{tabular}{ |c|c|c|c|c| } 
\hline
input size & Dijkstra & Bellman-Ford & Floyd-Warshall & QPU programming time \\
\hline
10 & 0.000587 & 0.000528 & 0.001080 & 0.010777  \\
25 & 0.001099 & 0.001113 & 0.013042 & 0.010791 \\
50 & 0.002357 & 0.002367 & 0.031585 & 0.010862 \\
100 & 0.004490 & 0.008783 & 0.216659 & 0.010898 \\
150 & 0.009010 & 0.011219 & 0.649454 & 0.011044 \\
200 & 0.010869 & 0.013536 & 1.503497 & 0.011038 \\
250 & 0.016921 & 0.017834 & 3.113248 & 0.011096 \\
300 & 0.018479 & 0.021513 & 5.324389 & 0.011044 \\
350 & 0.021103 & 0.021052 & 8.536259 & 0.011133 \\
400 & 0.025277 & 0.028834 & 11.65324 & 0.011181 \\
450 & 0.036367 & 0.036126 & 17.12447 & 0.011214 \\
500 & 0.034515 & 0.038908 & 23.60214 & 0.011213 \\
550 & 0.044816 & 0.045145 & 30.97148 & 0.011272 \\
600 & 0.052074 & 0.048889 & 41.73172 & 0.011249\\
\hline
\end{tabular}}
\caption{The table shows the run times for different shortest path algorithms in existence for simple graph with uniform edge weights.}
\end{table}
\begin{figure}[h!]
\centering
\includegraphics[width=1.1\linewidth]{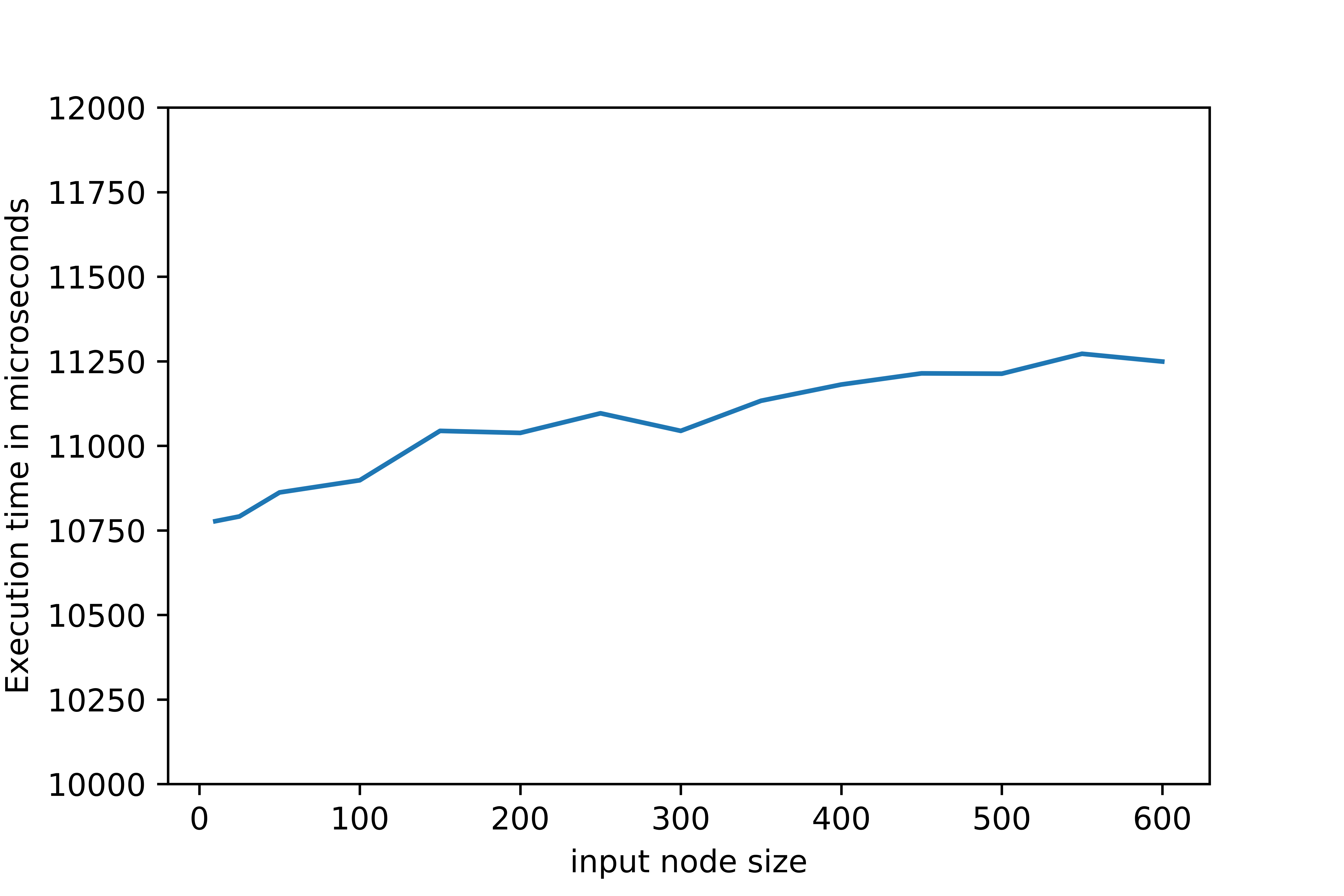}
\caption{Plot of programming time Vs. Input size on the D-wave QPU for table 3}
\label{"number of vertices vs time on the QPU"}
\end{figure}

\begin{figure}[h!]
\centering
\includegraphics[width=1.1\linewidth]{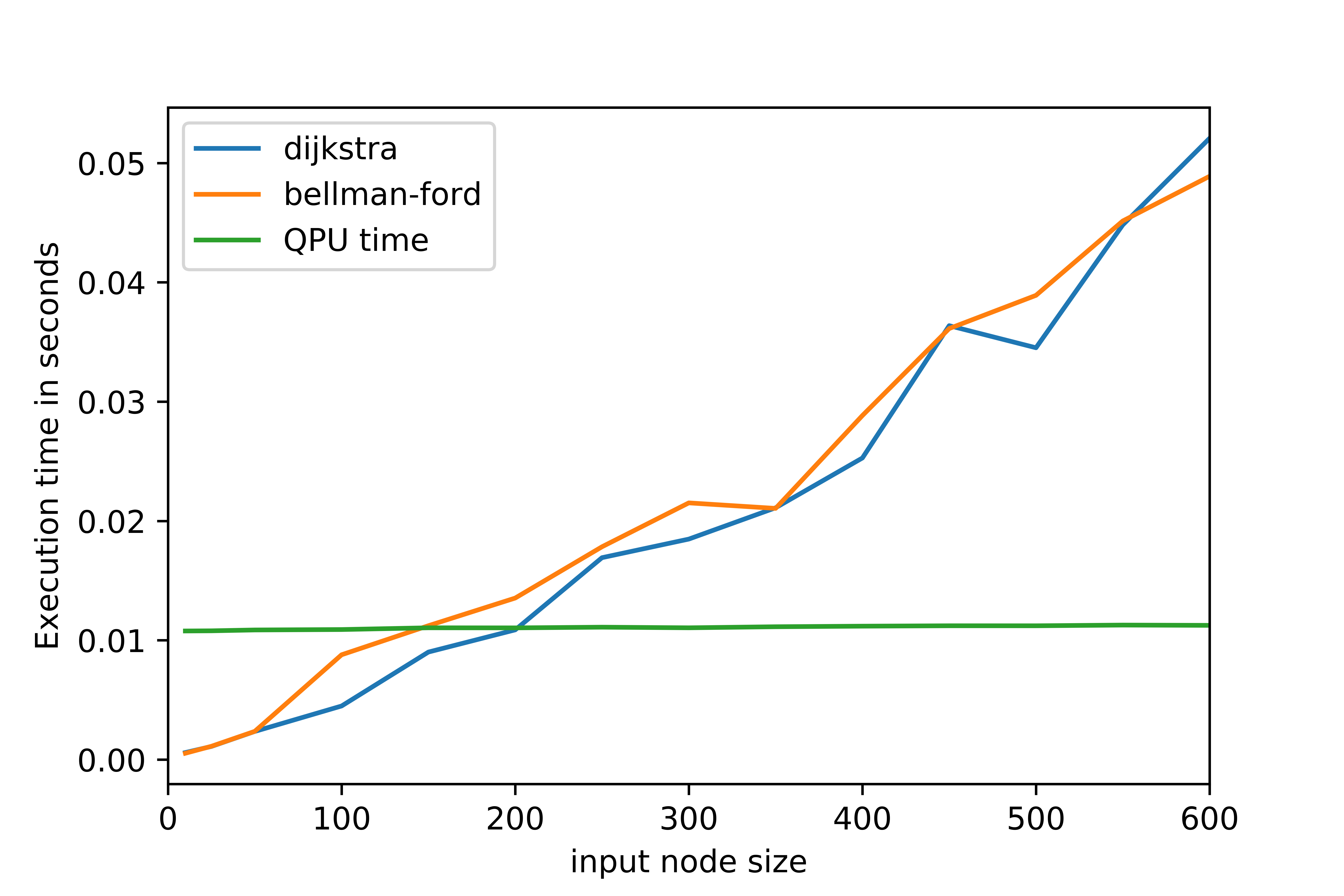}
\caption{Comparison of the different optimisation algorithms, estimating their run time Vs. Input size}
\label{"vertices vs time"}
\end{figure}
We have left out the Floyd-Warshall algorithm in our data plots due to its computational cost, which is much greater than the others considered here. Ideally the computation time taken to run the Dijkstra algorithm for small graphs would be in the order of nanoseconds, however using the built in shortest path algorithms from networkx yielded microsecond results. For this reason we have attempted to redo the experiment for comparative performance evaluation, we have re-plotted the results using fresh data from a new experiment. In order to determine the utility of quantum annealing to be of current practical use in attempting to solve the shortest path problem more efficiently as compared to the Dijkstra method, which also includes returning the Dijkstra path, we have used the most widely available open source Dijkstra algorithm as a metric for comparison instead of the networkx method. In order to get more insights into the quantum annealer, We have also manually recorded how long it takes to remotely send the problem over the cloud and get a solution returned. This is detailed in figure 20 as the overall execution time.

\begin{table}[ht!]
\centering
\scalebox{0.85}{
\begin{tabular}{ |c|c|c|c|c| } 
\hline
input size & Dijkstra time & overall execution time & QPU programming time \\
\hline
10&0.000275&0.01895&0.232645\\
20&0.000664&0.02992&0.232659\\
50&0.00198&0.13843&0.221894\\
100&0.00659&0.34757&0.232869\\
150&0.012168&0.8305&0.23288\\
200&0.020768&1.03941&0.232852\\
250&0.066757&1.84839&0.232911\\
300&0.086515&2.3749&0.233051\\
350&0.101589&3.56443&0.233064\\
400&0.092303&2.97474&0.233118\\
450&0.134814&5.69356&0.233159\\
500&0.193205&6.03239&0.233163\\
550&0.20353&8.27645&0.233164\\
600&0.402889&8.16396&0.23323\\
650&0.541983&5.59194&0.23318\\
700&0.560074&8.105979&0.233206\\
\hline
\end{tabular}}
\caption{The table shows the run times for running and displaying the shortest path on the D-wave processor and the native processor.}
\end{table}
\begin{figure}[h!]
\centering
\includegraphics[width=1.1\linewidth]{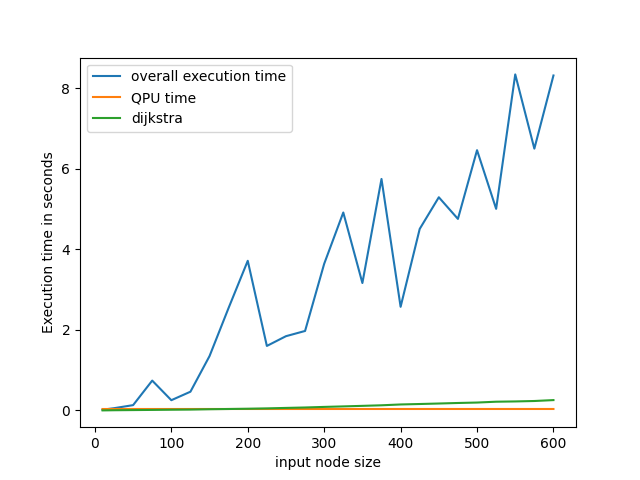}
\caption{Plot of programming time Vs. Input size which includes a manual time estimation to submit a problem and obtain a solution on the cloud.}
\label{"number of vertices vs time on the QPU"}
\end{figure}

\begin{figure}[h!]
\centering
\includegraphics[width=1.0\linewidth]{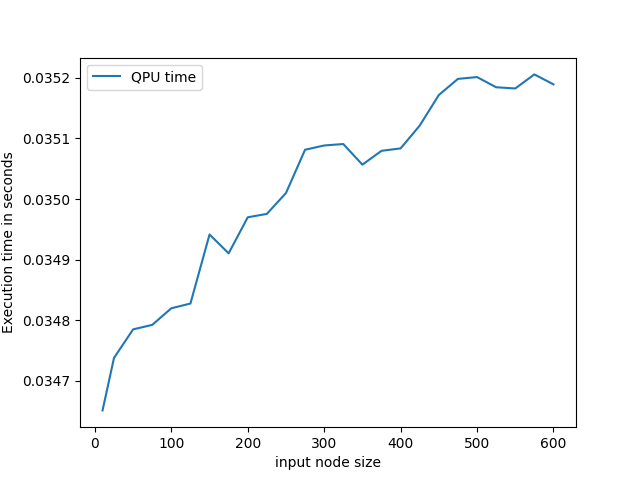}
\caption{Plot of programming time Vs. Input size on the D-wave QPU for table 4}
\label{"number of vertices vs time on the QPU"}
\end{figure}

\begin{figure}[h!]
\centering
\includegraphics[width=1.1\linewidth]{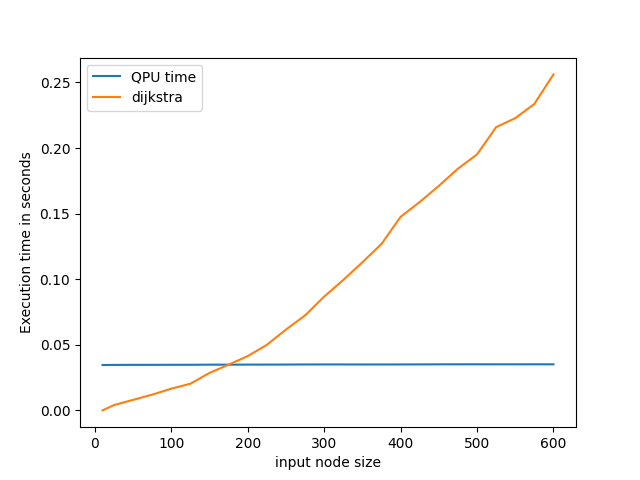}
\caption{Comparison of the Dijkstra Vs. the quantum annealing approach}
\label{"vertices vs time"}
\end{figure}

As such a quantum annealer with this degree of parallelism could be useful in modelling stochastic processes, Markov networks \cite{8} and other such models where there is a continuous change in the system parameters over time. There are many applications in the field of machine learning, since graph convolution on a network is an isomorphism that changes the structure of the graph over time \cite{9}. This mode of computation is best suited wherever dynamic optimization has to be performed ad-hoc, with the minimum time requirement in computing the optimization step.

After estimating the performance of the program with increasing number of vertices, the performance with respect to increase in the number of edges is recorded to finally estimate the time complexity of the algorithm. We used an edge connection probability based approach to populate the edges in the graph, and the run-time is recorded. The results were plotted with the edge population probabilities versus the run-time and curve fitting methods were used to estimate the best and the worst case complexity.

\begin{figure}[h!]
\centering
\includegraphics[width=1.1\linewidth]{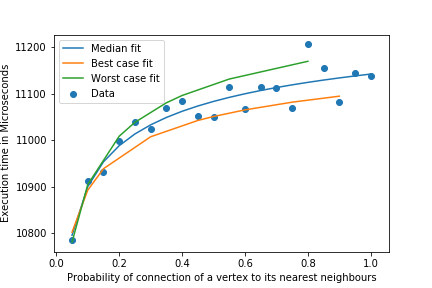}
\caption{Edge probabilities vs time on the D-wave QPU with a polynomial fit}
\label{"Edge probabilities vs time on the QPU with an polynomial fit"}
\end{figure}

\begin{figure}[h!]
\centering
\includegraphics[width=1.1\linewidth]{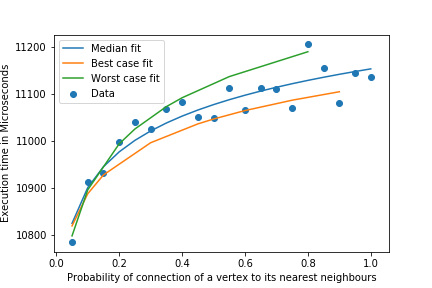}
\caption{Plot of programming time Vs. Edge probabilities on the D-wave QPU with a logarithmic fit}
\label{"vertices vs time"}
\end{figure}

After fitting the data, we get the expressions for the best case and the worst case time complexities. The best case gives us $O( V + 1/\sqrt{\Delta E})$
and the worst case gives us $O( V + log(\Delta E))$. Since not all vertices are connected in a random graph generator, the $\Delta E$ for the vertices in the expression for time complexity denotes the maximum degree of the vertex. It should be noted that the best case complexity term is just an ideal case scenario and is most highly unlikely to be the run-time complexity of this algorithm. We have taken into account the fact that, the polynomial fit tends to have a deeper curvature till it becomes somewhat constant as opposed to the logarithmic fit which doesn't seem to converge to a constant value. This explains why the lowest data points can be better represented by a polynomial fit. The logarithmic fit however is a more realistic estimate of the time complexity.

\bigskip

\begin{table}[ht!]
\centering
\begin{tabular}{p{4cm} | p{3.5cm}} 
\hline
Shortest Path Algorithm & Time Complexity \\
\hline
Bellman-Ford &  O($VE$) \\
Floyd-Warshall &  O($V^3$) \\
Dijkstra &  O($V^2$) \\
Dijkstra with Binary Heap & O($(E + V) log V$)\\
Dijkstra with Fibonacci Heap & O$(E + V log V)$ \\
Quantum annealing best case & O$(V + 1/\sqrt{\Delta E})$ \\
Quantum annealing worst case & O$(V + log(\Delta E))$ \\
\hline
\end{tabular}
\caption{A comparison of time complexities of the selected shortest path algorithms}
\end{table}

\bigskip

It has been observed that the AQC model is also suitable for heuristic algorithms where we start with a trial solution (ansatz) to a problem and slowly optimise this solution to meet the expected value of a known parameter of a system \cite{10}.

The results obtained were sourced from programs that we have created ourselves and are successfully scaled to make complete use of the QPU capabilities. Technologies such as AI \cite{11}, Internet of Things rely on graph structures and cost functions, which are traditionally optimisation problems.

 \section{Conclusion}
 In this paper we have managed to show that we can embed large, scale free graphs onto an adiabatic quantum computer to the maximum extent of its capabilities to solve binary optimisation problems. We were attempting to scale QUBO models which have been previously established in great detail in recent times \cite{12}. Since we have prioritised optimal performance in terms of speed and scale to account for impact, We have introduced constraints to the range of values the edge weights to simplify the problem so that on a large scale, QUBO models can be accurately run and can show significant speed up for a classic NP-hard problem. Further work can be done in embedding real world networks on many such devices, to solve network optimisation problems which are currently unsolvable by their classical counterparts. As adiabatic quantum computers improve in qubit sizes and the degree of connectivity of the qubits in the processor, the advantage of quantum computation over its classical counterpart can have a much greater impact in solving NP-hard combinatorial optimisation problems which are instrumental in many real world applications today. Although the QPU's continue to mature, graph embedding is an issue and we have not made any attempt to use an embedding algorithm apart from the custom embedding offered by Dwave. It is of great importance to create efficient embedding algorithms specific to quantum annealers in order to bridge the gap between mainstream optimization and quantum annealing.

 For the purpose of simulating real world networks with large sizes and high connectivity, efficient embedding programs on multiple D-wave processors are the logical next step in the endeavour to bring this mode of computing into mainstream applications \cite{13}. However with the release of the Pegasus chips, we expect to see QPUs with 5000 Qubit capacity, with 15 connections per Qubit \cite{14}. Such advancements enable more complex problems to be solved by quantum computers. A great reference for many of the concepts we have used in solving this problem is Network Science \cite{15} since AQC is a mode suited best for graph optimization problems.

\appendix
The degree distribution can be calculated by finding the probability that a vertex has a degree smaller than k, i.e $(k_i < k)$ and the vertices are added at a constant rate $P(t_i) = 1/{n_0 + t}$.

\begin{equation}
  \frac{dk_i}{dt} = np_i = n(k_i/\sum_{j}k_j)  
\end{equation}
The value for $\sum_{j}k_j$ is shown to be 2nt for undirected graphs. We now solve this differential equation to obtain a relation for $k_i$ with respect to time

\begin{equation}
  \frac{dk_i}{k_i} = (dt/2t)  
\end{equation}

\begin{equation}
  \log{k_i} = (1/2)\log{t}  
\end{equation}

\begin{equation}
  k_i = n(t/t_i)^{1/2}  
\end{equation}

After solving this differential equation we can now define the degree distribution as a cumulative distribution function $P(k_i)$.
The probability distribution for vertices with degrees k or smaller is given as 

\begin{equation}
  P(k_i < k) = [ \int_{0}^{k} p(k_i < k) \; dk ] 
\end{equation}
where $p(k_i < k)$ is the probability that a vertex has a degree smaller than k. Therefore, the above expression can be rewritten as

\begin{equation}
  p(k_i < k) = \frac{dP(k_i < k)}{dk} 
\end{equation}

From the calculations above we can now explicitly define a degree distribution function which is described below.

\begin{equation}
  P(t_i < nt/k^2) = P(k_i < k) = 1 - nt/{k^2}{P(t_i)}  
\end{equation}

\begin{equation}
 P(k_i < k) = 1 - (nt/{k^2})(1/{n_0 + t})  
\end{equation}
Differentiating this expression we get $p(k_i < k)$ which we then compare to the scaling relation $p(k_i < k) = k^{-\gamma}$ ,

\begin{equation}
  p(k_i < k) = \frac{d}{dk}P(k_i < k) = \frac{2n^2 t}{k^3} \frac{1}{(n_0 + t)}
\end{equation}
It is clear from this expression that $\gamma$ takes a value of 3. We have also verified the form of the equation by using curve fitting methods and the plot fits the curve for a polynomial fit of the form $1/(ax + b)^3$. The plot is presented below.

\begin{figure}[h!]
\centering
\includegraphics[width=1.1\linewidth]{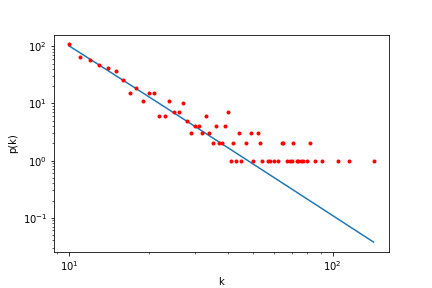}
\caption{The fitting of data points for a $\gamma$ value of 3 for the degree distribution plot of a Barabasi-Albert graph with 600 vertices and 10 incident edges for each incoming vertex i.e n = 10}
\label{"Degree distribution of the Barabasi-Albert network using a polynomial fit"}
\end{figure}

\section*{Acknowledgement}
We acknowledge the support by the U.S. Army under contact No. W15QKN-18-D-0040.

\end{document}